\documentclass[aps,pra,showpacs,twocolumn,nofootinbib]{revtex4-2}
\usepackage{bm}
\usepackage{amsmath}
\usepackage{amssymb}
\usepackage{slashed}
\usepackage{float}
\allowdisplaybreaks
\usepackage{subcaption}
\usepackage{dcolumn}
\usepackage{graphicx}

\usepackage{physics}
\usepackage{nicefrac}
\newcolumntype{w}[1]{D{.}{.}{#1}}
\newcommand{\Za}{Z\alpha}
\newcommand{\vare}{\varepsilon}
\newcommand{\balpha}{{\bm \alpha}}

\newcommand{\lbr}{\langle}
\newcommand{\rbr}{\rangle}

\newcommand{\rp}{{\rm p}}

\newcommand{\rk}{{\rm k}}
\newcommand{\rl}{{\rm l}}
\newcommand{\bfp}{{\bf p}}

\newcommand{\bfq}{{\bf q}}
\newcommand{\bfx}{{\bf x}}

\newcommand{\bfz}{{\bf z}}

\newcommand{\crossed}[1]{#1\!\!\!/}

\begin{document}

\title{Two-loop electron self-energy
with
accelerated partial-wave expansion
}

\author{V.~A. Yerokhin}
\email{Corresponding author: vladimir.yerokhin@mpi-hd.mpg.de}
\affiliation{Max~Planck~Institute for Nuclear Physics, Saupfercheckweg~1, D~69117 Heidelberg, Germany}

\author{Z. Harman}
\affiliation{Max~Planck~Institute for Nuclear Physics, Saupfercheckweg~1, D~69117 Heidelberg, Germany}

\author{C.~H. Keitel}
\affiliation{Max~Planck~Institute for Nuclear Physics, Saupfercheckweg~1, D~69117 Heidelberg, Germany}

\begin{abstract}

Calculations of the two-loop electron self-energy for the $n = 1$ and $n = 2$ states
of hydrogen-like ions are reported, performed to all orders in the nuclear binding strength parameter
$\Za$ (where $Z$ is the nuclear charge number and $\alpha$ is the fine structure constant).
The presented approach features an accelerated convergence of the partial-wave expansion
and allows calculations to be accomplished for nuclear charges lower  than previously possible
and with a higher numerical accuracy.

\end{abstract}

\maketitle

\section{Introduction}

\begin{figure}
\centerline{
\resizebox{\columnwidth}{!}{%
  \includegraphics{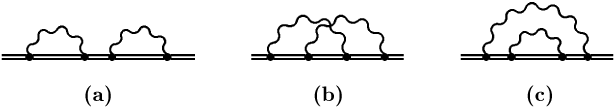}
}}
 \caption{
Feynman diagrams representing the two-loop electron self-energy:
(a) loop-after-loop, (b) overlapping, (c) nested diagram.
The double line denotes the electron in the
presence of the binding nuclear field; the wavy line denotes
a virtual photon.
\label{fig:sese}}
\end{figure}

The electron self-energy is the largest quantum electrodynamics (QED) effect
in atomic energy levels.
To match the precision of modern experiments
\cite{loetzsch:24,pfafflein:24,gassner:18,kraft:17},
this effect needs to be calculated
to all orders in the nuclear binding strength parameter
$\Za$ and with high accuracy \cite{indelicato:19}.
All-order calculations of the one-loop electron
self-energy started already in 1970's
\cite{mohr:74:a,mohr:74:b,mohr:82,mohr:92:b} and nowadays are performed routinely
\cite{yerokhin:25:se}.
By contrast,
calculations of the two-loop electron self-energy (SESE, Fig.~\ref{fig:sese})
turned out to be much more problematic and remained an open challenge for a long time.

Historically, the SESE correction was first studied within
the expansion in the parameter $\Za$.
Some of the expansion coefficients found in these calculations
turned out to be unexpectedly large and pushed theoretical predictions beyond their error
margins. So, in 1994,
Pachucki's calculation of the SESE correction of order $m\alpha^2(\Za)^5$
revealed \cite{pachucki:94} a large contribution that resolved
a disturbing discrepancy \cite{weitz:94} between theory and experiment
existing at that time.
Seven years later, another calculation by Pachucki
\cite{pachucki:01:pra}  identified a large SESE contribution of order
$m\alpha^2(\Za)^6\ln(\Za)^{-2}$,
once again correcting the previous theoretical prediction \cite{eides:01}.

In early 2000's, a break-through in two-loop calculations was achieved  \cite{yerokhin:01:sese}
and it became possible to compute the SESE effect to all orders in $\Za$.
As a result of many-year efforts, all-order SESE calculations were conducted
for the $n = 1$ and $n = 2$ states of hydrogen-like ions
\cite{yerokhin:03:prl,yerokhin:05:sese,yerokhin:06:prl,yerokhin:09:sese,yerokhin:18:sese}.
Their results
were successfully  validated by experiments in the high-$Z$ region
\cite{beiersdorfer:98,beiersdorfer:05}.
However, these calculations did not extend to the region of $Z < 10$
and the numerical accuracy in the lower-$Z$ range was rather limited.
In particular, for the experimentally important case of hydrogen,
direct all-order calculations were impossible and
an extrapolation from higher values of $Z$ was required.

Extrapolations of the all-order results
to $Z = 1$ reported in Refs.~\cite{yerokhin:05:sese,yerokhin:09:sese}
revealed a tension with predictions of the $\Za$ expansion
\cite{pachucki:03:prl}.
It was, however, argued \cite{karshenboim:19:sese,yerokhin:18:hydr} that this tension could
be plausibly explained by unknown higher-order $\Za$-expansion terms. So,
the recommended value for hydrogen \cite{tiesinga:21:codata18}
was obtained by assuming the consistency of the $\Za$-expansion
and the all-order results. The optimistic error obtained in this manner
nevertheless
constituted one of the two primary theoretical uncertainties in the hydrogen Lamb shift
\cite{tiesinga:21:codata18}.

The main factor limiting the accuracy of the previous all-order calculations
\cite{yerokhin:03:prl,yerokhin:05:sese,yerokhin:06:prl,yerokhin:09:sese,yerokhin:18:sese}
was the convergence of the partial-wave expansion of the electron propagators.
In the two-loop SESE diagrams there are 
two unbounded partial-wave expansions which
need to be extrapolated, with the number of
terms rapidly growing as $(2L)^3$ with increase of the cutoff parameter $L$.
This difficulty prevented further progress in extending calculations
in the low-$Z$ region.

Recently, advanced subtraction schemes were developed for the 
case of the one-loop self-energy
\cite{yerokhin:05:se,sapirstein:23},
which provided a significantly improved convergence of the partial-wave
expansion.
One of these schemes \cite{sapirstein:23} turned out to be
simple enough to allow generalizations
to higher orders of perturbation theory \cite{malyshev:24}.
In our recent Letter
\cite{yerokhin:24:sese}, we generalized this approach to the case of the
two-loop self-energy,
demonstrated a drastic improvement in the partial-wave convergence,
and performed calculations for the $1s$ state and $Z = 5$--$50$. As a result,
a $3.5\sigma$ disagreement was revealed between the extrapolated 
nonperturbative and $\Za$-expansion
results for hydrogen.
The resulting shift in the theoretical prediction for the $2s$-$1s$ transition frequency
in hydrogen impacted the determination of the Rydberg constant based
on this transition. Specifically, our calculation of the SESE correction
for hydrogen \cite{yerokhin:24:sese} decreased the value of
the Rydberg constant by $3.3$~kHz or 1.4$\sigma$ \cite{mohr:24:codata}.

The goal of the present work is to extend the generalization of the
two-loop
accelerated-convergence scheme to the excited states
and apply the developed method for extensive computations for
the $n = 1$ and $n = 2$ states.
The paper is organized as follows. In Sec.~\ref{sec:acc} we describe the
general idea of the convergence acceleration of the partial-wave
expansion. Sec.~\ref{sec:basic} summarizes basic formulas for the
SESE correction. Sec.~\ref{sec:M} describes the part of the SESE correction
calculated in the coordinate space, the so-called $M$ term. Sec.~\ref{sec:P}
discuss the part calculated in the mixed momentum-coordinate representation,
the so-called $P$ term. Sec.~\ref{sec:Fterm} describes the part
calculated in the momentum space, the $F$ term. In Sec.~\ref{sec:tot}
we collect all parts together and obtain the complete SESE correction.
Numerical results and discussion are presented in Sec.~\ref{sec:res}.

The relativistic units ($\hbar=c=m=1$) and the Heaviside charge units ($ \alpha = e^2/4\pi$, $e<0$)
are used throughout this paper.
We use roman style ($\rp$) for four vectors, boldface ($\bfp$) for three
vectors and italic style ($p$) for scalars.
Four vectors have the form $\rp = (\rp_0,\bfp)$.

\section{Convergence acceleration of partial-wave expansion}
\label{sec:acc}

The Dirac-Coulomb Green function $G$ can be represented as an expansion in the number of interactions
with the binding Coulomb field $V_C$
\begin{align}\label{eq:acc1}
G(\vare,\bfx_1,\bfx_2)  = &\ G^{(0)}(\vare,\bfx_1,\bfx_2)
 + G^{(1)}(\vare,\bfx_1,\bfx_2)
  \nonumber \\ &
 + G^{(2)}(\vare,\bfx_1,\bfx_2)  +
\ldots\,,
\end{align}
where $\vare$, $\bfx_1$, $\bfx_2$ are the energy and the two radial arguments, 
correspondingly, 
and the index $k$ in $G^{(k)}$ denotes the number of interactions with $V_C$.
The renormalization procedure of the electron self-energy typically involves
\cite{snyderman:91} separation of the two first terms of the above expansion,
$G^{(0)}$ and $G^{(1)}$, and their calculation in the momentum space. The third
term,  $G^{(2)}$, is usually not separated out since its calculation in the
momentum space would be too cumbersome to be practical. At the same time,
computing $G^{(2)}$ without any partial-wave expansion
would be advantageous since this term is usually responsible for the slowest-converging
part of the partial-wave expansion. The general idea of the
accelerating-convergence method \cite{sapirstein:23} is to
separate out a suitably chosen approximation for $G^{(2)}$,
which has a similar
partial-wave expansion but is more tractable in practical computations.

The basic idea goes back to the calculation of P.~Mohr \cite{mohr:74:a}, who
introduced the following approximation for the one-potential Green function
$G^{(1)}$ in the region $\bfx_1\approx\bfx_2$:
\begin{align}\label{eq:2}
G^{(1)}(\vare,\bfx_1,\bfx_2) &\ = \int d\bfz\, G^{(0)}(\vare,\bfx_1,\bfz)\,
   V_C(z)\, G^{(0)}(\vare,\bfz,\bfx_2)
  \nonumber \\ &
\approx  V_C(x_1)\, \int d\bfz\, G^{(0)}(\vare,\bfx_1,\bfz)\,
    G^{(0)}(\vare,\bfz,\bfx_2)
  \nonumber \\ &
=  V_C(x_1)\,\dot{G}^{(0)}(\vare,\bfx_1,\bfx_2)
\,,
\end{align}
where
\begin{align}
\dot{G}^{(0)}(\vare,\bfx_1,\bfx_2) = -\frac{\partial }{\partial \vare}\,G^{(0)}(\vare,\bfx_1,\bfx_2)\,.
\end{align}
The approximation (\ref{eq:2}) neglects the commutator $[G^{(0)},V_C]$, which is
small in the region $\bfx_1\approx\bfx_2$ and does not change significantly the
partial-wave expansion.

Sapirstein and Cheng \cite{sapirstein:23} used the same reasoning to obtain
an approximation for the two-potential Green function,
\begin{align}
G^{(2)}(\vare,\bfx_1,\bfx_2) &\
 \approx V_C(x_1)\,\ddot{G}^{(0)}(\vare,\bfx_1,\bfx_2)\, V_C(x_2)\,,
\end{align}
where
\begin{align}
\ddot{G}^{(0)}(\vare,\bfx_1,\bfx_2) =&\ \frac12\,\frac{\partial^2 }{(\partial \vare)^2}\,G^{(0)}(\vare,\bfx_1,\bfx_2)\,.
\end{align}
It is important that
$\dot{G}^{(0)}$ and $\ddot{G}^{(0)}$ are known both in coordinate space in the form of the partial-wave expansion as
well as in a closed form in momentum space.
Therefore, one can subtract expressions with $\dot{G}^{(0)}$ and $\ddot{G}^{(0)}$ in coordinate space
and then re-add them,
computed in the momentum space and without any partial-wave expansion.
Ref.~\cite{sapirstein:23} demonstrated that this approach yields a drastic improvement
in the convergence of the partial-wave expansion
for the one-loop self-energy. In this work we will generalize this
approach to the two-loop self-energy.

We also mention another accelerated-convergence approach developed in Ref.~\cite{yerokhin:05:se}.
It (approximately) accounts for not only the
two-potential Green function $G^{(2)}$ but also the three- and more-potential contributions
and yields typically an even better convergence acceleration (see the discussion in
Ref.~\cite{yerokhin:25:se}). However,
extending this  method to higher-order self-energy diagrams turned out to be problematic and
has never been demonstrated so far.

\section{Two-loop self-energy: general formulas}
\label{sec:basic}

The two-loop self-energy correction is depicted in Fig.~\ref{fig:sese} and
can be represented as a sum of four terms,
\begin{align}  \label{eq1}
E_{\rm SESE} = E_{\rm LAL}+ E_{\rm red} + E_{O} + E_{N} \,.
\end{align}
The first term, known as the loop-after-loop (LAL) correction, is
induced by the irreducible ($n\ne a$) part of the diagram in Fig.~\ref{fig:sese}(a).
It is given by
\begin{align}  \label{eq2}
E_{\rm LAL} =
\sum_{n\neq a} \frac1{\vare_a-\vare_n}
\bra{a} \Sigma(\vare_a) \ket{n}
\bra{n} \Sigma(\vare_a) \ket{a}\,,
\end{align}
where $\vare_a$ is the Dirac energy of the reference state $a$, the summation over $n$ runs
over the Dirac spectrum, and
$\Sigma(\vare)$ is the self-energy operator defined as
\begin{eqnarray}  \label{eq3}
\Sigma(\vare,\bfx_1,\bfx_2) &=& 2i\alpha \int_{C_F} d\omega\,
    D^{\mu\nu}(\omega,x_{12})\,
 \nonumber \\ && \times
      \alpha_{\mu}
         G(\vare-\omega,\bfx_1,\bfx_2) \alpha_{\nu}
         \,.
\end{eqnarray}
Here,
$D^{\mu\nu}(\omega,x)$ is the photon propagator,
$\alpha_{\mu} = (1,\balpha)$ are the Dirac matrices,
$G(\vare,\bfx_i,\bfx_j)$ is the
Dirac-Coulomb Green function,
$x_{ij} = |\bfx_i-\bfx_j|$, and $C_F$ is
the standard Feynman integration contour.
The LAL correction is can be renormalized and
calculated separately. It was computed already in
Refs.~\cite{mitrushenkov:95,mallampalli:98:prl,yerokhin:00:lalpra}
and thus will not be discussed here.

The second term in Eq.~(\ref{eq1}), denoted as the reducible contribution in the following, is
the reducible $(n=a)$ part of the diagram in Fig.~\ref{fig:sese}(a).
It is given by
\begin{equation}    \label{red}
E_{\rm red} =
\bra{a} \Sigma(\vare_a)\ket{a}\,
  \bra{a}
    \frac{\partial}{\partial \vare} \Sigma (\vare)
            \ket{a}
             \Big|_{\vare=\vare_a}
            \,.
\end{equation}

\begin{widetext}

The third term in Eq.~(\ref{eq1}) is induced by the {\em overlapping} diagram shown in Fig.~\ref{fig:sese}(b)
and is expressed as
\begin{align}  \label{eq5}
\Delta E_{O} =  &\, (2i\alpha)^2
   \int_{C_F} d\omega_1d\omega_2
   \int d\bfx_1 d\bfx_2 d\bfx_3 d\bfx_4\,
    D^{\mu\nu}(\omega_1,x_{13})\,
    D^{\rho\sigma}(\omega_2,x_{24})\,
\nonumber \\ &  \times
   {\psi}^{\dag}_a(\bfx_1)\, \alpha_{\mu}\, G(\vare_a-\omega_1,\bfx_1,\bfx_2)\,
           \alpha_{\rho}\,
  G(\vare_a-\omega_1-\omega_2,\bfx_2,\bfx_3)\,\alpha_{\nu}\, G(\vare_a-\omega_2,\bfx_3,\bfx_4)\,
         \alpha_{\sigma}
     \psi_a(\bfx_4)\,.
\end{align}

The {\em nested } diagram shown
in Fig.~\ref{fig:sese}(c) gives rise to the fourth term in Eq.~(\ref{eq1}),
expressed as
\begin{align}  \label{eq6}
\Delta E_{N} = &\, (2i\alpha)^2 \int_{C_F} d\omega_1
   \int d\bfx_1 d\bfx_2 d\bfx_3 d\bfx_4\,
    D^{\mu\nu}(\omega_1,x_{14})\,
\nonumber \\ &  \times
       {\psi}^{\dag}_a(\bfx_1)\, \alpha_{\mu}
          G(\vare_a-\omega_1,\bfx_1,\bfx_2) \,
    \Sigma(\vare_a-\omega_1,\bfx_2,\bfx_3)          \,
          G(\vare_a-\omega_1,\bfx_3,\bfx_4)\,
          \alpha_{\nu}\, \psi_a(\bfx_4)\,.
\end{align}

\end{widetext}

The above formulas are formal expressions that require renormalization before any actual calculations can be performed.
The renormalization scheme for the two-loop self-energy was formulated
in Ref.~\cite{mallampalli:98:pra} and fully implemented in a series of our studies \cite{yerokhin:01:sese,yerokhin:03:prl,yerokhin:06:prl}.
The scheme is based on subtracting and re-adding one or two first terms of the expansion of 
the Dirac-Coulomb propagators
in terms of interactions with the binding field, see Eq.~(\ref{eq1}).
The resulting diagrams to be computed are divided into three classes:
(i) those calculated in coordinate space ($M$-term),
(ii) those calculated in the mixed momentum-coordinate representation ($P$-term) and
(iii) those computed in momentum space ($F$-term).
In this way, the renormalized reducible, overlapping and nested contributions were divided into
the $M$, $P$, and $F$ parts, each of which are finite and can be evaluated numerically,
\begin{align}
E_{{\rm red},R} = E_{{\rm red},M}  + E_{{\rm red},F}\,, \\
E_{O,R} = E_{O,M} + 2\,E_{O,P} + E_{O,F}\,, \\
E_{N,R} = E_{N,M} + E_{N,P} + E_{N,F}\,,
\end{align}
where the subscript $R$ marks the renormalized contributions.
A detailed description of this separation will be provided in the following sections.

In the present work we will extend the previous renormalization scheme
for the overlapping and nested contributions, by subtracting and
re-adding additional
diagrams containing two Coulomb interactions inside the self-energy loops.
After this, the extended separation becomes
\begin{align}
E_{O,R} = E_{O,M'} + 2\,E_{O,P'} + E_{O,F'}\,, \\
E_{N,R} = E_{N,M'} + E_{N,P'} + E_{N,F'}\,,
\end{align}
as will be elaborated in the following sections. For the reducible term,
the acceleration of the partial-wave expansion will not be discussed here.
It can be accomplished in the same way as for the one-loop self-energy in Ref.~\cite{yerokhin:05:se}
and was implemented already in our previous investigations \cite{yerokhin:18:sese}.
As a result, $E_{{\rm red},M}$ is
computed to a very high accuracy
and does not contribute to the total uncertainty of the two-loop result.

\begin{figure*}
\centerline{
\resizebox{0.75\textwidth}{!}{%
  \includegraphics{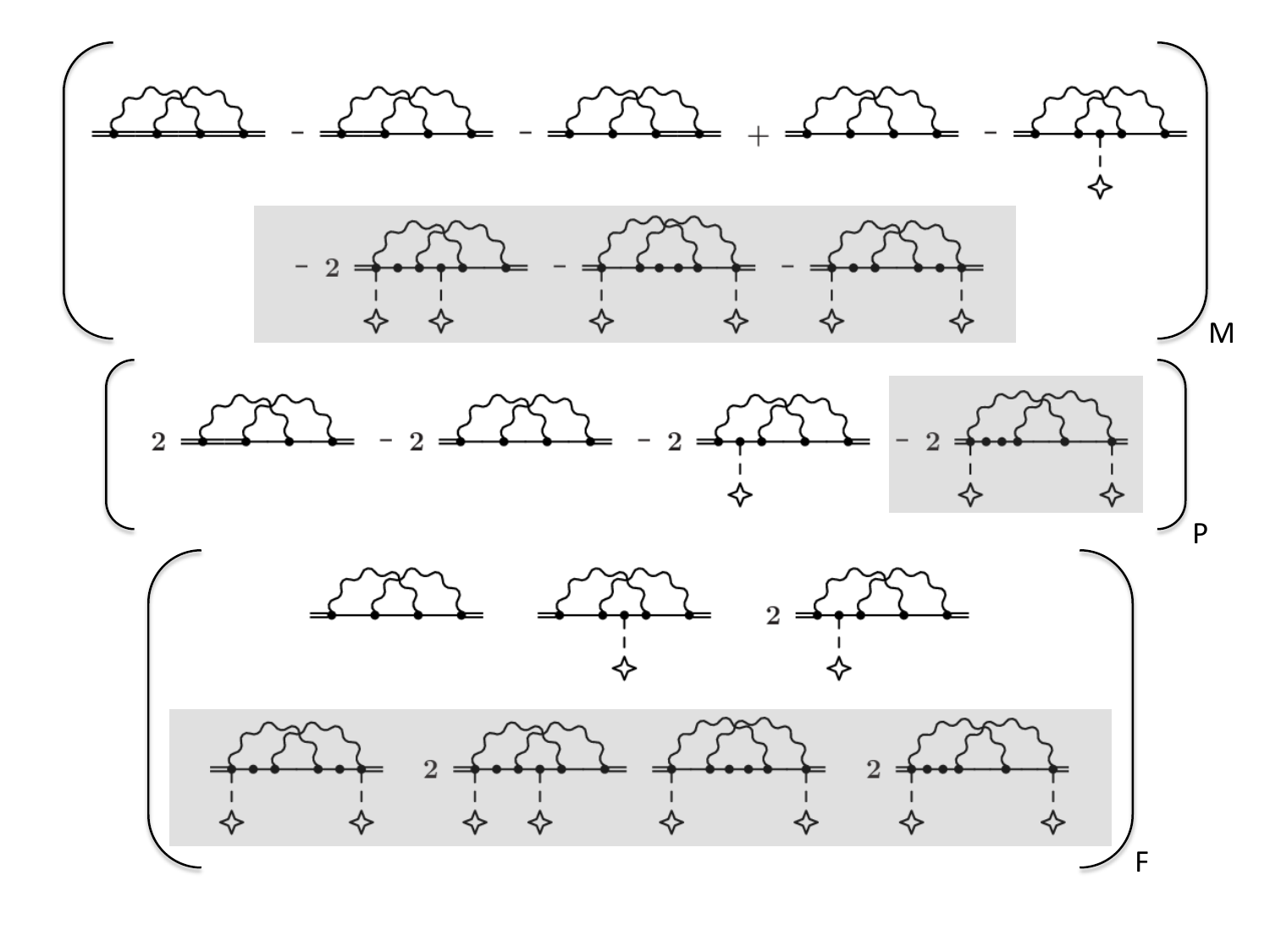}
}
}
 \caption{
Two subtraction schemes used for calculations of the
overlapping diagram: the standard scheme (without shaded diagrams) and
the accelerated-convergence scheme (with shaded
diagrams), see text.
Diagrams are divided into three groups, labelled as ``M'' (the $M$ term),
``P'' (the $P$ term), and ``F'' (the $F$ term), see text.
The single solid line represents the free-electron propagator,
while the dashed line terminated by a stylized cross indicates the
interaction with the binding nuclear field.
The reference-state infrared subtractions are not shown.
\label{fig:overlap}}
\end{figure*}

\begin{figure*}
\centerline{
\resizebox{0.75\textwidth}{!}{%
  \includegraphics{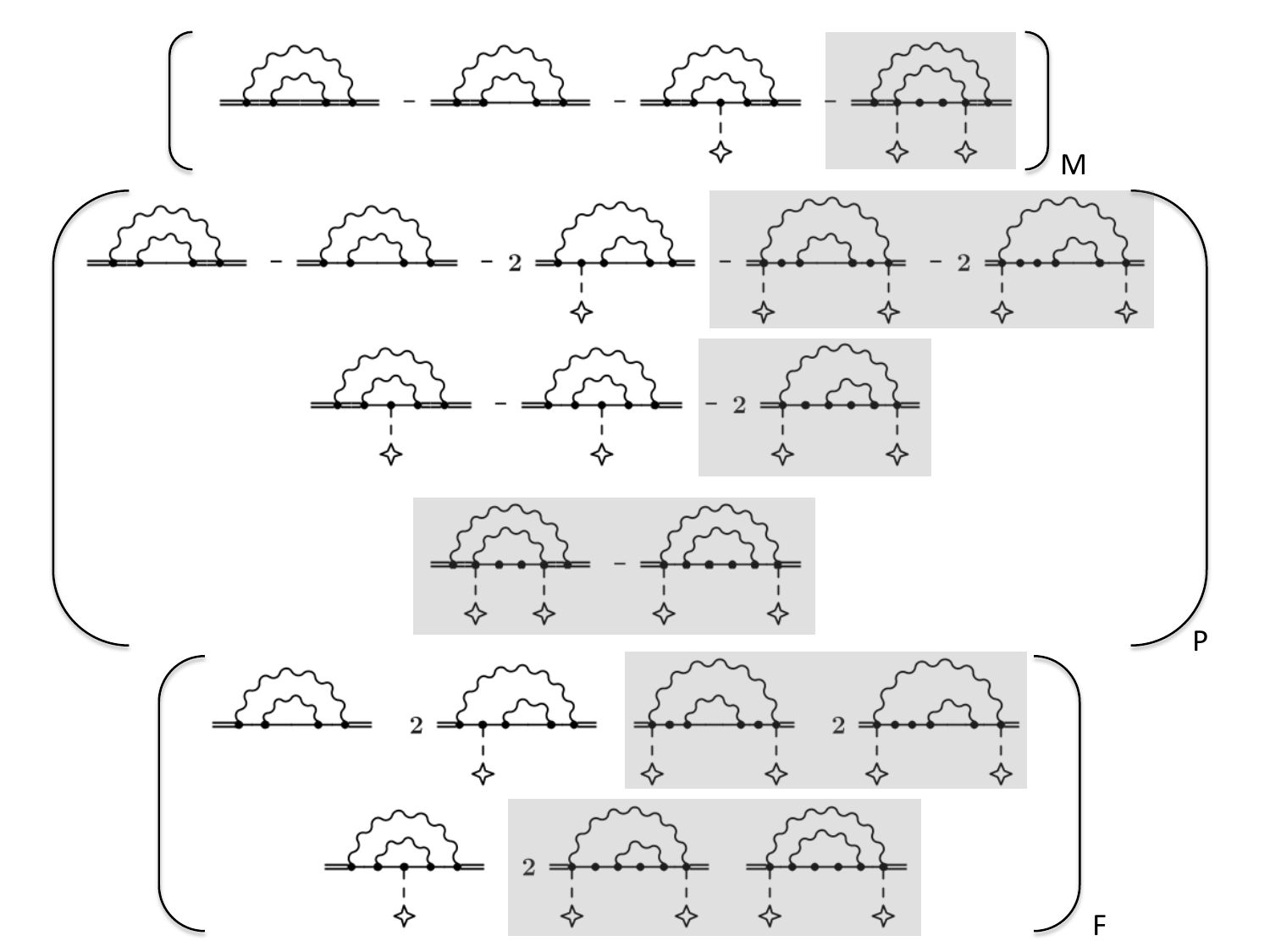}
}
}
 \caption{Same as Fig.~\ref{fig:overlap} for the
nested diagram. \label{fig:nested}}
\end{figure*}

\section{$\bm{M}$ term}
\label{sec:M}

The $M$-term contributions contain three Dirac-Coulomb electron propagators but 
are ultraviolet finite.
They are computed in the coordinate space and contain a double partial-wave expansion.
This is the most computationally intensive part of the calculation.
The $M$ term consists of the reducible, overlapping, and nested $M$ contributions.
They are obtained from the corresponding unrenormalized expressions by
subtracting from the Dirac-Coulomb propagators $G$ 
the corresponding contributions of the
free propagators $G^{(0)}$ and sometimes the one-potential
propagators $G^{(1)}$, as discussed below.
The accelerated-convergence scheme involves additional
subtractions of terms with two Coulomb interactions.

\begin{widetext}

\subsection{Reducible $\bm{M}$ term}

The reducible $M$ term is defined by (see Eq.~(25) of Ref.~\cite{yerokhin:18:sese})
\begin{align}    \label{redM}
E_{{\rm red},M} =
-2i\alpha\, E_{\rm SE}
 \int_{C_F} \!\!\! d\omega\,
  \int d\bfx_1\,d\bfx_2\,d\bfx_3\,
    D^{\mu\nu}(\omega,x_{13})\,
    \psi_a^{\dag}(\bfx_1)\,
      \alpha_{\mu}
      \Big[
         G(\vare_a-\omega,\bfx_1,\bfx_2)\,
         G(\vare_a-\omega,\bfx_2,\bfx_3)
         -\ldots
         \Big]
         \alpha_{\nu}\,
    \psi_a(\bfx_3)\,
            \,,
\end{align}
where $E_{\rm SE} = \bra{a} \Sigma_R(\vare_a)\ket{a}$ is the renormalized one-loop self-energy correction
and $\ldots$ denotes subtractions defined by
\begin{align}   \label{redMb}
G\,G \to G\,G - G^{(0)}\,G^{(0)} - G^{(a)}\,G^{(a)}\,.
\end{align}
Here,
$G^{(a)}(\vare,\bfx_1,\bfx_2) = \sum_{\mu_{a'}} \psi_{a'}(\bfx_1)\,\psi_{a'}^{\dag}(\bfx_2)/(\vare-\vare_a)$
is the reference-state part of the Dirac-Coulomb Green function, and
$a'$ are the electron states that differ from the reference state $a$ only by
the angular-momentum projection $\mu_{a'}$. The subtraction of the last term in Eq.~(\ref{redMb})
removes the
reference-state infrared divergency otherwise present at $\omega \to 0$.

The acceleration of the partial-wave expansion for $E_{{\rm red},M}$ is relatively straightforward
and will not be discussed here.
It was accomplished in the same way as for the one-loop self-energy in Ref.~\cite{yerokhin:05:se}
and implemented already in our previous investigations \cite{yerokhin:18:sese}.

\subsection{Overlapping $\bm{M}$ term}

The overlapping $M$ term $E_{O,M}$
is obtained from $E_{O}$ in Eq.~(\ref{eq5}) by
applying the following subtractions, see Eq.~(21) of Ref.~\cite{yerokhin:18:sese}
and the first line of Fig.~\ref{fig:overlap}:
\begin{align}  \label{eqM1}
 G\, G\, G \  & \to
   G\, G\, G
 - G\, G^{(0)}\,G^{(0)}
 - G^{(0)}\,G^{(0)}\,G
 + G ^{(0)}\,G ^{(0)}\,G ^{(0)}
 - G ^{(0)}\,G ^{(1)}\,G ^{(0)}\,.
\end{align}

In order to improve the convergence of the partial-wave expansion in $E_{O,M}$, we
introduce additional subtractions and then re-add the subtracted terms
calculated in a closed form in momentum space.
More specifically, we subtract and re-add the diagrams with two Coulomb interactions
inside the loops.
We thus represent $E_{O,M}$ as (see the second line of Fig.~\ref{fig:overlap})
\begin{align} \label{eqM1b}
E_{O,M} = E_{O,M'} + E_{o110} + E_{o011} + E_{o020} + E_{o101}\,,
\end{align}
where $E_{O,M'}$ contains in addition to (\ref{eqM1}) the following subtractions:
\begin{align}  \label{eqM2}
 G\, G\, G \  & \to \ldots
 - V_C\, \dot{G}^{(0)} \,G^{(1)} \,G ^{(0)}
 - {G}^{(0)} \,{G}^{(1)} \,\dot{G} ^{(0)}\,V_C
 - V_C\, G^{(0)} \,\ddot{G}^{(0)} \,G ^{(0)}\,V_C
 - V_C\, \dot{G}^{(0)} \,G^{(0)} \,\dot{G} ^{(0)}\,V_C\,.
\end{align}
The last four terms in Eq.~(\ref{eqM1b}) corresponds to the four subtracted terms in Eq.~(\ref{eqM2}).
In the notation $E_{oijk}$, $i$, $j$, and $k$ denote the number of Coulomb interactions
in the first, second, and third electron propagator, respectively.
In Eq.~(\ref{eqM2}) and in other shortened formulas below
we assume the implicit ordering of radial arguments, 
e.g., $V_C\,G \,G \,G \,V_C$ should be understood as
$V_C(x_1)\,G (\bfx_1,\bfx_2)\,G (\bfx_2,\bfx_3)\,G (\bfx_3,\bfx_4)\,V_C(x_4)$.

The last four terms on the right-hand-side of Eq.~(\ref{eqM1b}) 
contain only the free-electron propagators and
are calculated in momentum space, as described in Sec.~\ref{sec:Fterm},
after taking into account that $E_{o110} = E_{o011}$.
We note that in $E_{o110}$ and $E_{o011}$, unlike in all other subtraction terms,
we kept one Coulomb interaction intact (as $G^{(1)}$, rather than commuting it outside).
It was done because such a subtraction yielded a somewhat better partial-wave convergence.

\subsection{Nested $\bm{M}$ term}

The nested $M$ term $E_{N,M}$
is obtained from $E_{N}$ in Eq.~(\ref{eq6}) by applying the following subtractions,
see Eq.~(16) of Ref.~\cite{yerokhin:18:sese}
and the first line of Fig.~\ref{fig:nested}:
\begin{align}  \label{eqM4}
G \,\Sigma(\vare_a - \omega_1)\,G
 \to
G \,\Sigma^{(2+)}(\vare_a - \omega_1)\,G
-
G^{(a)} \,\Sigma^{(2+)}(\vare_a)\,G^{(a)}
\,,
\end{align}
where
\begin{align}
\Sigma^{(2+)}(\vare,\bfx_1,\bfx_2) = &\
2i\alpha \int_{C_F} d\omega\,
    D^{\mu\nu}(\omega,x_{12})\,
      \alpha_{\mu}\,
      {G}^{(2+)}(\vare-\omega,\bfx_1,\bfx_2)\,
         \alpha_{\nu} \,,
\end{align}
and $G^{(2+)} =  G -  G^{(0)} -
G^{(1)}$ is the Dirac-Coulomb Green function with two or more interactions
with the binding Coulomb field.
We note that the subtraction of the last term in the right-hand-side of Eq.~(\ref{eqM4}) removes
the reference-state infrared divergency present in $E_N$.

In order to improve the convergence of the partial-wave expansion in $E_{N,M}$, we
introduce an additional subtraction and then re-add the subtracted term back.
So, $E_{N,M}$ is represented as
\begin{align} \label{eqM6}
E_{N,M} = E_{N,M'} + E_{N3,P}\,,
\end{align}
where $E_{N,M'}$ is obtained from Eq.~(\ref{eq6})
by applying the following subtractions (see the first line of Fig.~\ref{fig:nested}):
\begin{align}  \label{eqM5}
G\,\Sigma(\vare_a - \omega_1)\,G
 \to &\
G\,\Sigma_s^{(2+)}(\vare_a - \omega_1)\,G
-
G^{(a)}\,\Sigma_s^{(2+)}(\vare_a)\,G^{(a)}
\,,
\end{align}
where
\begin{align}
\Sigma_s^{(2+)}(\vare,\bfx_1,\bfx_2) = &\
2i\alpha \int_{C_F} d\omega\,
    D^{\mu\nu}(\omega,x_{12})\,
      \alpha_{\mu}\,
      \Big[
      {G}^{(2+)}(\vare-\omega,\bfx_1,\bfx_2)\,
      - V_C(x_1)\,
         \ddot{G}^{(0)}(\vare-\omega,\bfx_1,\bfx_2)\, V_C(x_2)\,
         \Big]
         \alpha_{\nu} \,.
\end{align}
We note that this is the same subtraction that was used in the calculation
of the one-loop self-energy correction in Ref.~\cite{sapirstein:23}. The term
$E_{N3,P}$ in Eq.~(\ref{eqM6}) represents the subtracted term calculated
separately. It
is computed in the mixed momentum-coordinate representation, with
one partial-wave expansion instead of two in $E_{N,M}$.
The calculation of $E_{N3,P}$ is analogous to the $P$-term contributions
and will be described in the next Section.

\section{$\bm{P}$ term}
\label{sec:P}

The $P$ term contains both the Dirac-Coulomb propagators and ultraviolet-divergent
subgraphs, whose renormalization is performed in the momentum space. For this reason, the
numerical evaluation of the $P$ terms is carried out in the mixed momentum-coordinate
representation. It involves the Fourier transforms of the Dirac-Coulomb propagators
over one radial variable and a single partial-wave expansion.

\subsection{Overlapping $\bm{P}$ term}

The overlapping $P$ term $E_{O,P}$ is expressed as [see Eq.~(120) of Ref.~\cite{yerokhin:03:epjd}
and the third line of Fig.~\ref{fig:overlap}]
\begin{align}
E_{O,P} =&\ -2i\alpha \int_{C_F} d\omega\, \int
\frac{d\bfp_1}{(2\pi)^3}\,
        \frac{d\bfp_2}{(2\pi)^3}\,
        \int d\bfx_1 \,
                \frac{\exp(-i\bfq\cdot \bfx_1)}{\omega^2 -\bfq^2+ i0}\,
\psi_a^{\dag}(\bfx_1) \alpha_{\mu}\,
 G^{(2+)}(E,\bfx_1,\bfp_1)\,
        \gamma^0 \Gamma^{\mu}_R(E,\bfp_1;\vare_a,\bfp_2)\,
                        \psi_a(\bfp_2)
\,,
\end{align}
where $E = \vare_a-\omega$,
$\bfq = \bfp_1-\bfp_2$,
and $\Gamma^{\mu}_R$ is the renormalized one-loop
vertex operator in momentum space, see Appendix~A of Ref.~\cite{yerokhin:99:pra}
for definition and explicit representation.

In order to improve the partial-wave convergence, we subtract and then re-add an
approximation for the two-potential Green function, representing $E_{O,P}$ as
\begin{align}\label{eqP1}
E_{O,P} = E_{O,P'} + E_{o200}\,,
\end{align}
with (see the third line of Fig.~\ref{fig:overlap})
\begin{align}
E_{O,P'} = -2i\alpha \int_{C_F} d\omega\, \int
\frac{d\bfp_1}{(2\pi)^3}\,
        \frac{d\bfp_2}{(2\pi)^3}\,
        \int d\bfx_1 \,
&\
                \frac{\exp(-i\bfq\cdot \bfx_1)}{\omega^2 -\bfq^2+ i0}\,
\psi_a^{\dag}(\bfx_1) \alpha_{\mu}
\Big[
 G^{(2+)}(E,\bfx_1,\bfp_1)\,
        \gamma^0 \Gamma^{\mu}_R(E,\bfp_1;\vare_a,\bfp_2)\,\psi_a(\bfp_2)
\nonumber \\ &
-
V_C(x_1)\,\ddot{G}^{(0)}(E,\bfx_1,\bfp_1)\,
       \gamma^0 \Gamma^{\mu}_R (E,\bfp_1;\vare_a,\bfp_2)\,\psi_{Va}(\bfp_2)
\Big]
\,,
\end{align}
where $\psi_{Va}(\bfp)$ is the Fourier transform of the product $V_C$ and $\psi_a$,
\begin{equation} \label{psiVa}
\psi_{Va}(\bfp) = \int d^3 \bfx\,
        e^{-i\bfp\cdot\bfx}\, V_C(x)\, \psi_{a}(\bfx)\,,
\end{equation}
and $E_{o200}$ is the subtracted term calculated separately without partial-wave expansion.
It contains only free-electron propagators and
is calculated in momentum space, see Sec.~\ref{sec:Fterm}.

\subsection{Nested $\bm{P}$ terms}

In the standard scheme, there are two nested $P$ terms;
they are referred to as the first and second $P$ contributions,
respectively.
In the accelerated scheme, the third $P$ contribution
$E_{N3,P}$ appears, introduced in Eq.~(\ref{eqM6}).

The first nested $P$ contribution is given by [see Eq.~(113) of Ref.~\cite{yerokhin:03:epjd}
and the second line of Fig.~\ref{fig:nested}]
\begin{align}                                           \label{N1P}
E_{N1,P} =&\
        2i\alpha \int_{C_F} d\omega\,
                \int \frac{d\bfp}{(2\pi)^3}\,
                        \int d\bfx_1 d \bfx_2 \,
        D^{\mu\nu}(\omega,x_{12})\,
                 \psi^{\dag}_a(\bfx_1) \alpha_{\mu}
      \Bigl[G(E,\bfx_1,\bfp)\,
                    \gamma^0 \Sigma^{(0)}_R(E,\bfp)\,
                        G(E,\bfp,\bfx_2)
                         -\ldots \Bigr]
         \alpha_{\nu}        \psi_a(\bfx_2)\, ,
\end{align}
where $\Sigma^{(0)}_R$ is the renormalized self-energy operator
in the momentum space, see Appendix A of Ref.~\cite{yerokhin:99:pra} for
definition and explicit formulas, and
$\ldots$ denotes subtractions schematically represented as
\begin{eqnarray}    \label{subtrN1P}
G\, \gamma^0 \Sigma(E)\, G &\to&
       G\, \gamma^0 \Sigma(E) G
     - G^{(0)}\, \gamma^0 \Sigma(E)\, G^{(0)}
    - G^{(1)}\, \gamma^0 \Sigma(E)\, G^{(0)}
     - G^{(0)}\, \gamma^0 \Sigma(E)\, G^{(1)}
     - G^{(a)}\, \gamma^0 \Sigma(\vare_a)\, G^{(a)}
     \,.
\end{eqnarray}
The last term in the above removes the reference-state infrared divergence.
As a result, $E_{N1,P}$ is both ultraviolet and infrared finite.

We now improve the convergence of the partial-wave expansion in $E_{N1,P}$ by
introducing additional subtractions with two Coulomb interactions inside the
electron loops
and then re-adding the subtracted contributions
calculated separately,
\begin{align}          \label{N1Pb}
E_{N1,P} = E_{N1,P'} + E_{n101} + E_{n200} + E_{n002}\,,
\end{align}
where $E_{N1,P'}$ is defined by
the subtractions (\ref{subtrN1P}) supplemented with
(see the second line of Fig.~\ref{fig:nested})
\begin{eqnarray}    \label{N1Pc}
G\, \gamma^0 \Sigma(E)\, G &\to&
 \ldots
 - V_C\, \dot{G}^{(0)}\, \gamma^0 \Sigma(E)\, \dot{G}^{(0)}\,V_C
 - V_C\, \ddot{G}^{(0)}\, \gamma^0 \Sigma(E)\, {G}^{(0)}\,V_C
 - V_C\, {G}^{(0)}\, \gamma^0 \Sigma(E)\, \ddot{G}^{(0)}\,V_C\,,
\end{eqnarray}
and the last three terms in the right-hand-side of Eq.~(\ref{N1Pb})
correspond to the three subtraction terms in Eq.~(\ref{N1Pc}).
We note that the symmetry relation ensures that $E_{n200} = E_{n002}$.
The subtracted terms contain only free-electron propagators and
are calculated in momentum space in Sec.~\ref{sec:Fterm}.

The second nested $P$ contribution $E_{N2,P}$ is given by
[see Eq.~(117) of Ref.~\cite{yerokhin:03:epjd} and
the third line of Fig.~\ref{fig:nested}]
\begin{align}                                           \label{N2P}
E_{N2,P} = &\
        2i\alpha \int_{C_F} d\omega\,
                \int \frac{d\bfp_1}{(2\pi)^3}\, \frac{d\bfp_2}{(2\pi)^3}\,
                        \int d\bfx_1 d \bfx_2 \,
        D^{\mu\nu}(\omega,x_{12})\, V_C(q)\,
                 \psi^{\dag}_a(\bfx_1)\,
                \nonumber \\ & \times
  \alpha_{\mu}
      \Bigl[G(E,\bfx_1,\bfp_1)\,
                    \gamma^0  \Gamma^0_R(E,\bfp_1;E,\bfp_2)\,
                        G(E,\bfp_2,\bfx_2)
                     -\ldots \Bigr]
         \alpha_{\nu}        \psi_a(\bfx_2)
         \, ,
\end{align}
where 
$\Gamma^{0}_R$ is the time ($\mu = 0$) component of the renormalized one-loop
vertex operator in momentum space
and $\ldots$ denotes subtractions defined as
\begin{equation}    \label{subtrN2P}
G\,\gamma^0 \Gamma(E;E)\, G
\to
G \gamma^0 \Gamma(E;E)\, G
- G^{(0)} \gamma^0 \Gamma(E;E)\, G^{(0)}
- G^{(a)} \gamma^0 \Gamma(\vare_a;\vare_a)\, G^{(a)}
\,.
\end{equation}
The last term in the above removes the reference-state infrared divergency.
As a result, $E_{N2,P}$ is both ultraviolet and infrared finite.

We now improve the convergence of the partial-wave expansion in $E_{N2,P}$ by
introducing additional subtractions,
\begin{align}       \label{eqN2Pb}
E_{N2,P} = E_{N2,P'} +  E_{n110}+  E_{n011}\,,
\end{align}
where $E_{N2,P'}$ is defined by
the subtractions (\ref{subtrN2P}) supplemented with
(see third line of Fig.~\ref{fig:nested})
\begin{eqnarray}
G\,\gamma^0 \Gamma(E;E)\, G
&\to&
 \ldots
 - V_C\, \dot{G}^{(0)}\, \gamma^0 \dot{\Sigma}(E)\, {G}^{(0)}\,V_C
 - V_C\, {G}^{(0)}\, \gamma^0 \dot{\Sigma}(E)\, \dot{G}^{(0)}\,V_C
 \,.
\end{eqnarray}
We note that the symmetry relation ensures that $E_{n110} = E_{n011}$.
The subtracted terms contain only free-electron propagators and
are calculated in momentum space in Sec.~\ref{sec:Fterm}.

Finally, we have to account for the $E_{N3,P}$ contribution in Eq.~(\ref{eqM6}).
For its evaluation we also introduce a subtraction,
\begin{align} \label{eqN3Pb}
E_{N3,P} = E_{N3,P'} + E_{n020}\,,
\end{align}
where
(see fourth line of Fig.~\ref{fig:nested})
\begin{align}                                           \label{N3P}
E_{N3,P} =&\
        2i\alpha \int_{C_F} d\omega\,
                \int \frac{d\bfp}{(2\pi)^3}\,
                        \int d\bfx_1 d \bfx_2 \,
        D^{\mu\nu}(\omega,x_{12})\,
                 \psi^{\dag}_a(\bfx_1) \alpha_{\mu}
      \Bigl[G_V(E,\bfx_1,\bfp)\,
                    \gamma^0 \ddot\Sigma^{(0)}_R(E,\bfp)\,
                        G_V(E,\bfp,\bfx_2)
                         - \ldots
                         \Bigr]
         \alpha_{\nu}        \psi_a(\bfx_2)\, ,
\end{align}
where
$\ldots$ denotes the subtraction
\begin{align}
G_V\,\gamma^0 \ddot\Sigma^{(0)}(E)\,G_V
\to
G_V\,\gamma^0 \ddot\Sigma^{(0)}(E)\,G_V
-
G^{(a)}_V\,\gamma^0 \ddot\Sigma^{(0)}(\vare_a)\,G^{(a)}_V
-
V_C\,G^{(0)}\,\gamma^0 \ddot\Sigma^{(0)}(E)\,G^{(0)}\,V_C
\,,
\end{align}
and
$G_V$ denotes the Fourier transform of the product of $G$ and $V_C$,
\begin{align} \label{eq3a1}
G_V(\vare,\bfx_1,\bfp) = &\
        \int d\bfx_2 \, e^{i\bfp\cdot \bfx_2} \,
       G(\vare,\bfx_1,\bfx_2) \, V_C(\bfx_2)
         \,,
\\
G_V(\vare,\bfp,\bfx_2) = &\
        \int d\bfx_1 \, e^{-i\bfp\cdot \bfx_1}\,
    V_C(\bfx_1)\,
          G(\vare,\bfx_1,\bfx_2) \, .
\end{align}
The subtraction term $E_{n020}$ contains only free-electron propagators and
is calculated in momentum space in the next Section.

\section{$\bm{F}$ term}
\label{sec:Fterm}

The definition of the $F$ term $E_F$ in the standard scheme
is rather long and is described in detail in Sec.~4 of
Ref.~\cite{yerokhin:03:epjd}. We here discuss the modification of the $F$ term
required in
the accelerated-convergence scheme. In this case the $F$ term
receives additional contributions
from the subtraction terms introduced in previous sections.
So, we define $E_{F'} = E_F + E_{F, \rm add}$, where
\begin{align} \label{eqF0}
E_{F, \rm add} = 2\,E_{o110} + E_{o020} + E_{o101} + 2\,E_{o200} + 2\,E_{n110} + 2\,E_{n200} + E_{n020} + E_{n101}\,.
\end{align}
All contributions to $E_{F, \rm add}$ are finite and can be evaluated in $D=4$ dimensions, which greatly simplifies
the derivation.

All contributions in the right-hand-side of Eq.~(\ref{eqF0}) except $E_{o110}$ have the structure
similar to that for the zero-potential $F$ term, namely,
\begin{align} \label{eqF1}
E_{i} = &\ \int \frac{d^3\bfp}{(2\pi)^3}\, \overline{\psi}_{Va}(\bfp)\, \Sigma_{i}(\rp)\,\psi_{Va}(\bfp)\,,
\end{align}
where $\overline{\psi} = \psi^{\dag}\gamma^0$,  $\psi_{Va}(\bfp)$ is defined by Eq.~(\ref{psiVa}), and
operators $\Sigma_i(\rp)$ are defined below. They represent the second derivative over the time component of
the incoming momentum of the zero-potential nested and overlapping operators, which appeared in
Ref.~\cite{yerokhin:03:epjd}.  Specifically,
the nested operators $\Sigma_i(\rp)$ are given by
\begin{align} \label{eqF5}
\Sigma_{n200}(\rp) = \frac{\alpha}{4\pi}\,
 \int \frac{d^4k}{i\pi^2}\,
   \frac1{\rk^2}\,
   \gamma_{\mu}
   \left[
      \frac12 \frac{\partial^2}{\partial \rp_0^2}
   \frac1{\crossed{\rp}-\crossed{\rk}-m}
   \right]
   \,
   \Sigma_R^{(0)}(\rp-\rk)\,
       \frac1{\crossed{\rp}-\crossed{\rk}-m}
    \gamma^{\mu}\,,
\end{align}
\begin{align} \label{eqF6}
\Sigma_{n110}(\rp) = \frac{\alpha}{4\pi}
 \int \frac{d^4k}{i\pi^2}\,
   \frac1{\rk^2}\,
   \gamma_{\mu}
   \left[
      \frac{\partial}{\partial \rp_0}
   \frac1{\crossed{\rp}-\crossed{\rk}-m}
   \right]
   \,
   \left[
      \frac{\partial}{\partial \rp_0}\,
   \Sigma_R^{(0)}(\rp-\rk)\,
   \right]
       \frac1{\crossed{\rp}-\crossed{\rk}-m}
    \gamma^{\mu}\,,
\end{align}
\begin{align} \label{eqF7}
\Sigma_{n020}(\rp) = \frac{\alpha}{4\pi}
 \int \frac{d^4k}{i\pi^2}\,
   \frac1{\rk^2}\,
   \gamma_{\mu}
   \frac1{\crossed{\rp}-\crossed{\rk}-m}
   \,
   \left[
    \frac12  \frac{\partial^2}{\partial \rp_0^2}\,
   \Sigma_R^{(0)}(\rp-\rk)\,
   \right]
       \frac1{\crossed{\rp}-\crossed{\rk}-m}
    \gamma^{\mu}\,,
\end{align}
\begin{align} \label{eqF8}
\Sigma_{n101}(\rp) = \frac{\alpha}{4\pi}\,
 \int \frac{d^4k}{i\pi^2}\,
   \frac1{\rk^2}\,
   \gamma_{\mu}
   \left[
      \frac{\partial}{\partial \rp_0}
   \frac1{\crossed{\rp}-\crossed{\rk}-m}
   \right]
   \,
   \Sigma_R^{(0)}(\rp-\rk)\,
   \left[
      \frac{\partial}{\partial \rp_0}
   \frac1{\crossed{\rp}-\crossed{\rk}-m}
   \right]
    \gamma^{\mu}\,,
\end{align}
where $\rp_0$ is the time component of the 4-vector $\rp = (\rp_0,\bfp)$,
derivatives are supposed to act only within the brackets
and $\Sigma_R^{(0)}$ is the renormalized free self-energy operator
in $D = 4$ dimensions, see Appendix A.1 of Ref.~\cite{yerokhin:03:epjd}
for the definition and the explicit representation.
The overlapping operators $\Sigma_i(\rp)$ are given by
\begin{align} \label{eqF2}
\Sigma_{o200}(\rp) = \frac{\alpha}{4\pi}\,
 \int \frac{d^4k}{i\pi^2}\,
   \frac1{\rk^2}\, \gamma_{\mu}
\left[
    \frac12 \frac{\partial^2}{\partial \rp_0^2}
   \frac1{\crossed{\rp}-\crossed{\rk}-m}\,
\right]\,
   \Gamma_R^{\mu}(\rp-\rk,\rp)\,,
\end{align}
\begin{align} \label{eqF3}
\Sigma_{o020}(\rp) = \left(\frac{\alpha}{4\pi}\right)^2\,
 \int \frac{d^4k}{i\pi^2}\,
    \int \frac{d^4l}{i\pi^2}
   \frac1{\rk^2\,\rl^2}\,
   \gamma_{\mu}
   \frac1{\crossed{\rp}-\crossed{\rk}-m}
   \,
    \gamma_{\nu} \,
\left[
    \frac12 \frac{\partial^2}{\partial \rp_0^2}
    \frac1{\crossed{\rp}-\crossed{\rk}-\crossed{\rl}-m}\,
\right]
    \gamma^{\mu}\,
       \frac1{\crossed{\rp}-\crossed{\rl}-m}
    \gamma^{\nu}\,,
\end{align}
\begin{align} \label{eqF4}
\Sigma_{o101}(\rp) = \left(\frac{\alpha}{4\pi}\right)^2\,
 \int \frac{d^4k}{i\pi^2}\,
    \int \frac{d^4l}{i\pi^2}
   \frac1{\rk^2\,\rl^2}\,
   \gamma_{\mu}
   \left[
      \frac{\partial}{\partial \rp_0}
   \frac1{\crossed{\rp}-\crossed{\rk}-m}
   \right]
   \,
    \gamma_{\nu} \,
    \frac1{\crossed{\rp}-\crossed{\rk}-\crossed{\rl}-m}\,
    \gamma^{\mu}\,
   \left[
      \frac{\partial}{\partial \rp_0}
       \frac1{\crossed{\rp}-\crossed{\rl}-m}
   \right]
    \gamma^{\nu}\,,
\end{align}
where
$\Gamma_R^{\mu}(\rp_1,\rp_2)$ is the renormalized one-loop vertex operator
in $D=4$ dimensions,
see Appendix A.2 of Ref.~\cite{yerokhin:03:epjd} for the definition and
explicit representation.

The $E_{o110}$ contribution has a different structure which is analogous to that for the one-potential
$F$ term,
\begin{align} \label{eqF10}
E_{o110} = &\ \int \frac{d^3\bfp_1}{(2\pi)^3}\,
            \frac{d^3\bfp_2}{(2\pi)^3}\,
\overline{\psi}_{Va}(\bfp_1)\,
  V_C(q)\, \Sigma_{o110}(\rp_1,\rp_2)\,\psi_{a}(\bfp_2)\,,
\end{align}
where $\rp_{1,0} = \rp_{2,0} = \vare_a$ and
\begin{align} \label{eqF11}
\Sigma_{o110}(\rp_1,\rp_2) =
\left(\frac{\alpha}{4\pi}\right)^2\,
 \int \frac{d^4k}{i\pi^2}\,
    \int \frac{d^4l}{i\pi^2}
   \frac1{\rk^2\,\rl^2}\,
   \gamma_{\mu}
   \left[
      -\frac{\partial}{\partial \rp_{1,0}}
   \frac1{\crossed{\rp}_1-\crossed{\rk}-m}
   \right]
   \,
    \gamma_{\nu} \,
    \frac1{\crossed{\rp}_1-\crossed{\rk}-\crossed{\rl}-m}\,
    \gamma_0
    \frac1{\crossed{\rp}_2-\crossed{\rk}-\crossed{\rl}-m}\,
    \gamma^{\mu}\,
       \frac1{\crossed{\rp}_2-\crossed{\rl}-m}
    \gamma^{\nu}\,.
\end{align}
Evaluation of the additional $F$-term contributions and their numerical computation
was carried out in full analogy with the calculation
of other $F$-term contributions  described
in Ref.~\cite{yerokhin:03:epjd}.

\end{widetext}

\section{Total two-loop self-energy}
\label{sec:tot}

Collecting all contributions discussed above, we write
the total two-loop self-energy correction as
\begin{align}
E_{\rm SESE} = &\ E_{\rm LAL} +
    \Big( E_{N,M'} + E_{O,M'} + E_{{\rm red},M} \Big)_{M'}
 \nonumber \\ &
   + \Big( E_{N1,P'} + E_{N2,P'} + E_{N3,P'} + 2\,E_{O,P'} \Big)_{P'}
 \nonumber \\ &
   + \Big( E_{F} + E_{F,\,{\rm add}}\Big)_{F'}
 \nonumber \\ &
 \equiv
 E_{\rm LAL} + E_{M'} + E_{P'} + E_{F'}\,.
\end{align}
The relation between $E_{M'}$, $E_{P'}$, $E_{F'}$ and the definitions used in our
previous studies is detailed in Appendix.

Numerical results for the two-loop self-energy correction
are conveniently parameterized in terms of the dimensionless function $F_{\rm SESE}$,
\begin{align}\label{eq:Za1}
E_{\rm SESE} = \left(\frac{\alpha}{\pi}\right)^2\, \frac{(\Za)^4}{n^3}\,
    F_{\rm SESE}(\Za)\,,
\end{align}
where $n$ is the principal quantum number of the reference state.
Historically, the  function $F_{\rm SESE}$ was investigated within the approach
based on the expansion in the parameter $\Za$. In order to compare our all-order
results with those previous studies, below we discuss the present status of the
$\Za$-expansion calculations of the SESE correction.

The $\Za$ expansion of the function $F_{\rm SESE}$ has the following form
\begin{align} \label{eq:Za2}
F_{\rm SESE}(\Za) =& \  B_{40} + (\Za)\,B_{50} + (\Za)^2\bigl[ B_{63}\, L^3
  \nonumber \\ &
+ B_{62}\, L^2 + B_{61}\, L + G_{60}(\Za)\bigr]\,,
\end{align}
where $L = \ln(\Za)^{-2}$ and
$G_{\rm 60}(\Za)$ is the remainder function containing higher-order expansion
terms in $\Za$. The $\Za$-expansion coefficients in Eq.~(\ref{eq:Za2}) are known
and  summarized in Table~3 of Ref.~\cite{yerokhin:18:hydr}.

The form of the $\Za$ expansion of the higher-order remainder $G_{60}(\Za)$
was worked out in Ref.~\cite{karshenboim:19:sese}
and is given by
\begin{align}\label{eq:Za3}
 G_{60}(\Za) = &\ B_{60} + (\Za)\big[ B_{72} \, L^2+  B_{71} \, L + B_{70}\big]
  \nonumber \\ &
  + (\Za)^2 \big[ B_{84}\,L^4 + B_{83}\,L^3 + B_{82}\,L^2
    \nonumber \\ &
\hspace*{2cm}
    + B_{81}\,L
+ B_{80}\big] + \ldots\,.
\end{align}
The coefficient $B_{60}$
was partially computed in Refs.~\cite{pachucki:03:prl,jentschura:05:sese,jentschura:06:sese}
to yield
\begin{align}\label{eq:Za4}
& B_{60}(1s) =  -61.6\,(9.2)\,,\nonumber\\
& B_{60}(2s) =  -53.2\,(8.0)\,,\nonumber\\
& B_{60}(2p_{1/2}) =  -1.5\,(3)\,,\nonumber\\
& B_{60}(2p_{3/2}) =  -1.8\,(3)\,,\nonumber\\
& B_{60}(3d_{3/2}) =   0.141\,(2)\,,\nonumber\\
& B_{60}(3d_{5/2}) =   0.123\,(2)\,,
\end{align}
where the uncertainty represents the estimation of unevaluated contributions.
The following differences of the $B_{60}$ coefficients were evaluated
completely \cite{jentschura:05:sese},
\begin{align}\label{eq:Za5}
& B_{60}(2s) - B_{60}(1s) =  14.1\,(4)\,,\nonumber\\
& B_{60}(2p_{3/2}) - B_{60}(2p_{1/2}) =  -0.361\,196\,,\nonumber\\
& B_{60}(3d_{5/2})-B_{60}(3d_{3/2}) = -0.018\,955\,.
\end{align}
The leading logarithmic coefficients in Eq.~(\ref{eq:Za3}) are known exactly
\cite{karshenboim:18,karshenboim:19:sese},
\begin{align}\label{eq:Za6}
B_{72} =&\  -\pi\Big( \frac{139}{48}- \frac43\,\ln 2\Big)\, \delta_{l,0}\,,
 \\
B_{84} = &\ -\frac7{27}\,\delta_{l,0}\,.
\end{align}
The next-to-leading logarithmic coefficient is known for states with $l>0$
and for the normalized difference of $S$ states \cite{karshenboim:18},
\begin{align}\label{eq:Za9}
& B_{71}(np) =\pi \Big( \frac{139}{144} - \frac{4}{9}\,\ln 2 \Big)
 \Big( 1-\frac1{n^2}\Big) \,, \\
& B_{71}(ns)- B_{71}(1s) =  \pi\, \Big( \frac{139}{12} -\frac{16}{3}\ln 2\Big) \,
  \Big[ \frac34 -\frac1n
  \nonumber \\ &
  + \frac1{4n^2} + \psi(n) + \gamma_E -\ln n\Big] \,,
\label{eq:Za10}
\end{align}
and $B_{71}(nl) = 0$ for states with $l > 1$. For the
specific case of $n = 2$, $B_{71}(2s)- B_{71}(1s) = 15.34528\ldots$.

\section{Numerical results}
\label{sec:res}

The accelerated-convergence scheme developed in this work involves subtracting and
re-adding a number of carefully selected subtraction terms, as compared to the
standard scheme used in the previous studies.
As a result, the subtraction terms have to be computed twice in two different ways.
The corresponding numerical results were shown to agree within the estimated
uncertainties, which provided an important check of consistency of our
numerical approach. Table~\ref{tab:subtr} presents an example
comparison for the subtraction terms computed in two ways.
Values obtained without partial-wave expansion (``0 PWE'') are contributions
to the $F$ term, computed in the momentum representation. Values labeled as
``1 PWE'' are computed in the mixed momentum-coordinate representation,
as the $P$-term contributions. Values labeled (``2 PWE'') are computed
in the coordinate representation, as the $M$-term contributions.
Apart from the consistency check,
the table demonstrates how reducing the number of partial-wave expansions
leads to significant improvements of numerical accuracy.

\begin{figure*}
\centerline{
\resizebox{0.9\textwidth}{!}{%
  \includegraphics{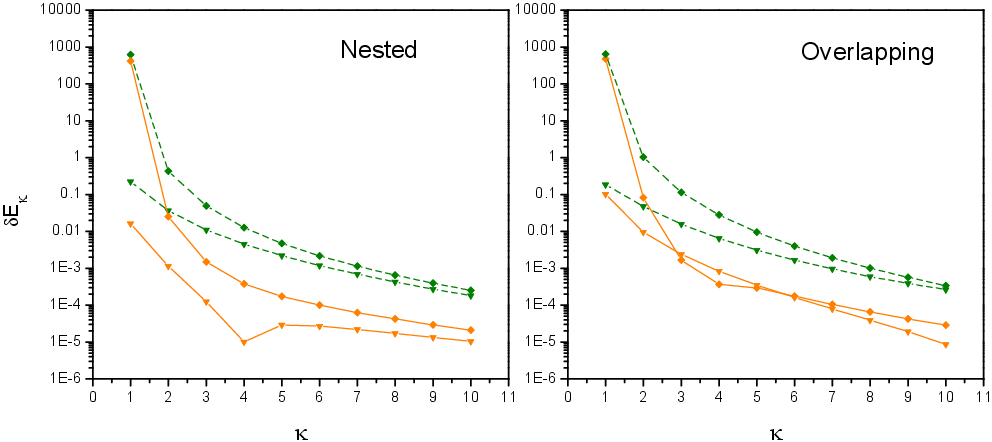}
}}
 \caption{
Comparison of convergence of the partial-wave expansion in the accelerated
(solid orange line) and the standard (dashed green line) approaches,
for the nested (left) and overlapping (right) $M$-term contributions,
for $Z = 10$ and the $1s$ state.
Plotted are the partial-wave expansion contributions
$\delta E_{\kappa_1,\kappa_2}$ for $(\kappa_1,\kappa_2) = (-\kappa,-\kappa)$
(diamonds) and $(\kappa_1,\kappa_2) = (\kappa,\kappa)$ (triangles),
in units of the function $F_{\rm SESE}(\Za)$ given by Eq.~(\ref{eq:Za1}).
\label{fig:mterm_special}}
\end{figure*}

Fig.~\ref{fig:mterm_special} illustrates the improvement of
convergence of the partial-wave expansion of the nested and overlapping
$M$-term contributions achieved
in the accelerated scheme as compared to the standard approach. We observe
that the additional subtractions introduced in the
accelerated scheme decrease the absolute values of higher-order
partial-wave expansion contributions by more than an order of magnitude.

\begin{table*}
 \caption{
Individual subtraction terms in the accelerated-convergence scheme,
calculated in different ways, for the $1s$ and $2p_{3/2}$ states and $Z = 83$,
 in units of $F_{\rm SESE}(\Za)$ given by Eq.~(\ref{eq:Za1}).
 Computations are carried out: (i) in the momentum space without
 partial-wave expansion (0 PWE), (ii) in the mixed momentum-coordinate representation
 with a single partial-wave expansion
 (1 PWE), (iii) in the coordinate space with double partial-wave expansion (2 PWE).
    \label{tab:subtr}}
\begin{ruledtabular}
    \begin{tabular}{l w{2.8} w{2.8} w{1.8} w{2.8} w{3.8} w{3.8} w{3.8} w{2.8} w{2.8}}

    &
 \multicolumn{1}{c}{$E_{n110}$} &
    \multicolumn{1}{c}{$E_{n200}$} &
        \multicolumn{1}{c}{$E_{n101}$} &
              \multicolumn{1}{c}{$E_{n020}$}  &
                  \multicolumn{1}{c}{$E_{N3P}$} &
 \multicolumn{1}{c}{$E_{o200}$} &
    \multicolumn{1}{c}{$E_{o110}$} &
        \multicolumn{1}{c}{$E_{o020}$} &
              \multicolumn{1}{c}{$E_{o101}$}
    \\
\hline\\[-5pt]
 \multicolumn{3}{l}{$1s$:} \\
%
%
0 PWE    & 0.083827       &  -0.509215      &  1.547290      &  -0.351166       &                    & -0.11774\,(1)   &  -0.322344      &   0.260970      &  -0.458625 \\
1 PWE   & 0.08384\,(2) &  -0.5092\,(1) &  1.5472\,(2) &  -0.35116\,(1) &  -0.16227\,(2) & -0.1178\,(1)    & \\
2 PWE   &                  &                   &                  &                    &  -0.1622\,(2)  &                     &  -0.3223\,(1) &   0.2608\,(5) &  -0.4586\,(8) \\[5pt]
%
%
%
 \multicolumn{3}{l}{$2p_{3/2}$:} \\
0 PWE   & -1.025996    & 1.821003     & 2.719870     & -0.576840     &               & -1.520245      & -0.650728    &  -0.144999     & -1.165321    \\
1 PWE   & -1.0260\,(1) & 1.8216\,(10) & 2.7189\,(17) & -0.57682\,(3) & -0.65471      & -1.5203\,(5)   &              &                               \\
2 PWE   &              &              &              &               & -0.6543\,(7)  &                & -0.6494\,(12)&  -0.1455\,(31) & -1.1646\,(56) \\

    \end{tabular}
\end{ruledtabular}
\end{table*}

\begin{table*}
 \caption{
 Individual contributions to the SESE correction for the $1s$, $2s$, $2p_{1/2}$, and $2p_{3/2}$
 states of hydrogen-like ions, in units of $F_{\rm SESE}(\Za)$ defined by Eq.~(\ref{eq:Za1}),
 and the higher-order remainder $G_{60}$ defined by Eq.~(\ref{eq:Za2}).
    \label{tab:numres}}
\begin{ruledtabular}
    \begin{tabular}{l w{1.5} w{3.6} w{3.6} w{3.6} w{1.6} w{3.8} }
 \multicolumn{1}{l}{$Z$}   & \multicolumn{1}{c}{$  E_{\rm LAL}$}  & \multicolumn{1}{c}{$  E_{F'}$}  & \multicolumn{1}{c}{$  E_{P'}$}  & \multicolumn{1}{c}{$  E_{M'}$}  & \multicolumn{1}{c}{$E_{\rm SESE}$}  & \multicolumn{1}{c}{$G_{60}(\Za)$} \\
\hline\\[-5pt]
$1s$ \\
  5   &        -0.1797         &      2692.9637\,(8)    &     -3650.2352\,(19)   &       958.1178\,(66)   &         0.6665\,(70)   &       -99.21\,(523)    \\
  6   &        -0.2188         &      1764.5418\,(4)    &     -2339.5596\,(14)   &       575.7840\,(33)   &         0.5474\,(36)   &      -100.51\,(190)    \\
  7   &        -0.2561         &      1230.6001\,(3)    &     -1598.7423\,(19)   &       368.8356\,(20)   &         0.4374\,(28)   &      -100.28\,(107)    \\
  8   &        -0.2917         &       898.4086\,(1)    &     -1145.2229\,(8)    &       247.4402\,(12)   &         0.3342\,(15)   &       -99.74\,(43)     \\
  9   &        -0.3255         &       679.2679\,(1)    &      -850.5560\,(6)    &       171.8521\,(11)   &         0.2385\,(13)   &       -98.66\,(29)     \\
 10   &        -0.3577         &       527.9948\,(1)    &      -650.0604\,(7)    &       122.5719\,(9)    &         0.1487\,(12)   &       -97.47\,(22)     \\
 12   &        -0.4170         &       339.8997\,(1)    &      -405.5446\,(9)    &        66.0473\,(7)    &        -0.0146\,(11)   &       -94.85\,(14)     \\
 14   &        -0.4703         &       233.0572\,(3)    &      -270.1819\,(7)    &        37.4359\,(6)    &        -0.1592\,(10)   &       -92.076\,(95)    \\
 16   &        -0.5184         &       167.3698\,(5)    &      -188.9373\,(3)    &        21.7979\,(7)    &        -0.2880\,(10)   &       -89.241\,(70)    \\
 18   &        -0.5620         &       124.5323\,(3)    &      -137.1426\,(7)    &        12.7680\,(6)    &        -0.4043\,(10)   &       -86.459\,(56)    \\
 20   &        -0.6015         &        95.2908\,(2)    &      -102.5403\,(6)    &         7.3419\,(4)    &        -0.5092\,(7)    &       -83.682\,(35)    \\
 22   &        -0.6376         &        74.5893\,(4)    &       -78.5427\,(1)    &         3.9864\,(6)    &        -0.6046\,(7)    &       -80.947\,(28)    \\
 26   &        -0.7014         &        48.2029\,(1)    &       -48.7936\,(4)    &         0.5211\,(6)    &        -0.7709\,(7)    &       -75.577\,(20)    \\
 30   &        -0.7565         &        32.8854         &       -32.1500\,(4)    &        -0.8930\,(8)    &        -0.9140\,(9)    &       -70.434\,(18)    \\
 34   &        -0.8053         &        23.3623         &       -22.1539\,(5)    &        -1.4414\,(8)    &        -1.0384\,(9)    &       -65.467\,(15)    \\
 40   &        -0.8711         &        14.8016         &       -13.5137\,(6)    &        -1.6182\,(5)    &        -1.2014\,(8)    &       -58.3741\,(91)   \\
 50   &        -0.9734         &         7.6735         &        -6.7227\,(4)    &        -1.4156\,(3)    &        -1.4382\,(5)    &       -47.4279\,(39)   \\
 60   &        -1.0825         &         4.3092         &        -3.7475\,(5)    &        -1.1460\,(4)    &        -1.6667\,(7)    &       -37.4734\,(35)   \\
 70   &        -1.2161         &         2.5100         &        -2.2797\,(7)    &        -0.9363\,(8)    &        -1.9220\,(11)   &       -28.4012\,(41)   \\
 83   &        -1.4658         &         1.2056         &        -1.3414\,(8)    &        -0.7597\,(5)    &        -2.3613\,(9)    &       -17.7985\,(25)   \\
 92   &        -1.7341\,(1)    &         0.6263         &        -1.0111\,(5)    &        -0.6894\,(9)    &        -2.8083\,(10)   &       -11.2195\,(23)   \\
100   &        -2.0989\,(1)    &         0.2105\,(1)    &        -0.8500\,(6)    &        -0.6571\,(7)    &        -3.3956\,(9)    &        -5.9350\,(17)   \\
\hline\\[-5pt]
$2s$ \\
 20   &        -0.3937\,(2)    &       231.6806\,(19)   &      -209.3077\,(34)   &       -22.4571\,(31)   &        -0.4779\,(50)   &       -69.83\,(23)     \\
 30   &        -0.4650\,(1)    &        88.1716\,(11)   &       -69.2962\,(14)   &       -19.3325\,(31)   &        -0.9221\,(35)   &       -58.113\,(74)    \\
 40   &        -0.5155\,(1)    &        43.9967\,(2)    &       -31.0717\,(11)   &       -13.6878\,(35)   &        -1.2784\,(37)   &       -47.594\,(44)    \\
 50   &        -0.5695         &        25.5056\,(1)    &       -16.7631\,(11)   &        -9.7890\,(25)   &        -1.6160\,(27)   &       -38.206\,(21)    \\
 60   &        -0.6434         &        16.2574\,(1)    &       -10.3822\,(13)   &        -7.2182\,(23)   &        -1.9863\,(26)   &       -29.829\,(14)    \\
 70   &        -0.7539         &        11.0465\,(1)    &        -7.2437\,(16)   &        -5.4923\,(20)   &        -2.4434\,(25)   &       -22.3699\,(97)   \\
 83   &        -0.9956         &         7.1031         &        -5.3773\,(21)   &        -4.0261\,(29)   &        -3.2959\,(36)   &       -13.9762\,(97)   \\
 92   &        -1.2839\,(1)    &         5.3296\,(1)    &        -4.9081\,(41)   &        -3.3484\,(39)   &        -4.2108\,(56)   &        -9.089\,(12)    \\
100   &        -1.7071\,(2)    &         4.0905\,(2)    &        -4.9248\,(41)   &        -2.9136\,(44)   &        -5.4551\,(60)   &        -5.541\,(11)    \\
\hline\\[-5pt]
$2p_{1/2}$ \\
 20   &         0.0288         &       258.3719\,(7)    &      -279.0481\,(18)   &        20.8021\,(63)   &         0.1547\,(66)   &        -0.55\,(31)     \\
 30   &         0.0255         &        96.6626\,(3)    &       -93.7366\,(11)   &        -2.7796\,(40)   &         0.1718\,(42)   &        -0.448\,(87)    \\
 40   &         0.0061         &        47.0988         &       -41.2883\,(5)    &        -5.6301\,(29)   &         0.1865\,(29)   &        -0.329\,(34)    \\
 50   &        -0.0294         &        26.5890         &       -21.0257\,(10)   &        -5.3356\,(29)   &         0.1983\,(31)   &        -0.232\,(23)    \\
 60   &        -0.0814         &        16.5294         &       -11.6820\,(15)   &        -4.5639\,(20)   &         0.2022\,(25)   &        -0.181\,(13)    \\
 70   &        -0.1524         &        11.0315         &        -6.8589\,(10)   &        -3.8279\,(13)   &         0.1922\,(17)   &        -0.1751\,(64)   \\
 83   &        -0.2869         &         7.1031         &        -3.6204\,(13)   &        -3.0621\,(10)   &         0.1336\,(17)   &        -0.2630\,(46)   \\
 92   &        -0.4343         &         5.5099         &        -2.3969\,(9)    &        -2.6606\,(13)   &         0.0181\,(16)   &        -0.4460\,(35)   \\
100   &        -0.6512         &         4.5466         &        -1.7174\,(8)    &        -2.3900\,(10)   &        -0.2120\,(13)   &        -0.7851\,(24)   \\
\hline\\[-5pt]
$2p_{3/2}$ \\
 30   &         0.0560         &        95.0135\,(3)    &       -95.0028\,(20)   &        -0.119\,(11)    &        -0.052\,(12)    &        -0.77\,(24)     \\
 40   &         0.0819         &        45.8258         &       -42.1612\,(9)    &        -3.8275\,(43)   &        -0.0810\,(44)   &        -0.834\,(51)    \\
 50   &         0.1105         &        25.5498         &       -21.7850\,(8)    &        -3.9872\,(26)   &        -0.1119\,(28)   &        -0.765\,(21)    \\
 60   &         0.1411         &        15.6315         &       -12.4081\,(16)   &        -3.5142\,(24)   &        -0.1498\,(29)   &        -0.706\,(15)    \\
 70   &         0.1723         &        10.2086         &        -7.5759\,(16)   &        -2.9998\,(21)   &        -0.1947\,(27)   &        -0.663\,(10)    \\
 83   &         0.2110         &         6.3020         &        -4.3263\,(16)   &        -2.4417\,(10)   &        -0.2551\,(19)   &        -0.6027\,(52)   \\
 92   &         0.2333         &         4.6820         &        -3.0789\,(21)   &        -2.1327\,(14)   &        -0.2964\,(25)   &        -0.5627\,(55)   \\
100   &         0.2469         &         3.6699         &        -2.3477\,(26)   &        -1.8972\,(33)   &        -0.3282\,(42)   &        -0.5225\,(78)
    \end{tabular}
\end{ruledtabular}
\end{table*}

Our numerical results
for the SESE correction
are presented in Table~\ref{tab:numres}
for the $1s$, $2s$, $2p_{1/2}$ and $2p_{3/2}$ states
of hydrogen-like ions.
The results are obtained for the point-charge nuclear model,
with the accelerated-convergence scheme developed in
this work. The obtained results
agree with those from our previous calculations
\cite{yerokhin:06:prl,yerokhin:09:sese,yerokhin:18:sese}
but are  more accurate, especially in the low-$Z$ region.
Numerical results for the $1s$ state and $Z\le 50$
were presented already in
our Letter \cite{yerokhin:24:sese}; here we extend these calculations to higher values of $Z$.
For the excited $n = 2$ states, we improve the numerical accuracy and extend the
calculated $Z$ region as compared to our previous calculations.

We now turn to the comparison of our nonperturbative results with calculations
performed
within the $\Za$-expansion. For the $1s$ state, a detailed analysis was already
presented in our Letter \cite{yerokhin:24:sese} and will not be repeated here. We recall
that it revealed a significant (3.5$\sigma$) deviation from results obtained
by $\Za$-expansion calculations \cite{pachucki:03:prl,karshenboim:19:sese}.

In the present work we performed calculations both for the $1s$ and $2s$ states
and thus have a possibility to study the normalized difference $\Delta_2 \equiv
8 E_{2s} - E_{1s}$, which is known
within the $\Za$ expansion to a greater extent than for
the $1s$ and $2s$ states
separately. In order to eliminate the rapidly varying structure at $Z\to 0$,
we define the nonlogarithmic higher-order remainder $ G_{60,\rm nlog}$ as
\begin{align}\label{eq:Gnlog}
 G_{60,\rm nlog}(\Za) =  G_{60}(\Za) - (\Za)\big[ B_{72} \, L^2+  B_{71} \, L \big]
 \,.
\end{align}
Using the fact that for the normalized $2s$-$1s$ difference and for the $2p_j$ states
the coefficient $B_{72}$ vanishes and
$B_{71}$  is known, see Eqs.~(\ref{eq:Za9}) and (\ref{eq:Za10}), we extract the numerical
values of $G_{60,\rm nlog}$
from our all-order numerical results.

\begin{figure}
\centerline{
\resizebox{\columnwidth}{!}{%
  \includegraphics{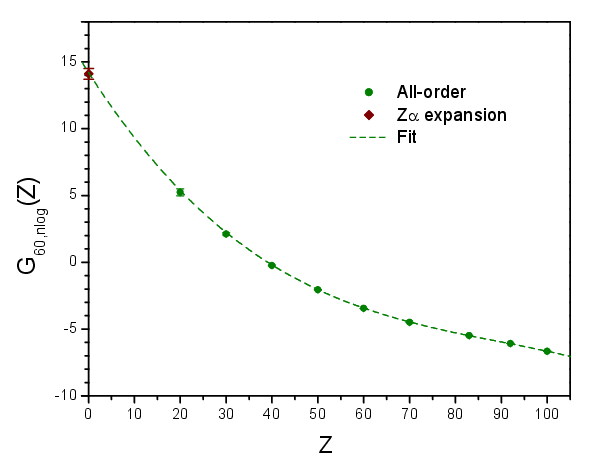}
}}
 \caption{The nonlogarithmic higher-order remainder for the normalized difference of the
 $2s$ and $1s$ states,
 $G_{60,\rm nlog}(2s\mbox{\rm-}1s) = G_{60,\rm nlog}(2s)-G_{60,\rm nlog}(1s)$,
 see Eq.~(\ref{eq:Gnlog}) for definition.
 Green dots
 denote our all-order numerical results; the brown dot
 at $Z = 0$ denotes the $\Za$-expansion limiting value,
 the dashed line is a polynomial fit.
 \label{fig:2s1s}}
\end{figure}

In Fig.~\ref{fig:2s1s} we present our nonperturbative values for the difference
$G_{60,\rm nlog}(2s)-G_{60,\rm nlog}(1s)$, plotted as a function of the
nuclear charge number $Z$, together with the $\Za$-expansion result for
the $B_{60}$ coefficient, which is the limiting value at $Z = 0$.
We observe that the nonperturbative values are consistent with
the $\Za$-expansion prediction, which is in contrast to the disagreement
observed for the $1s$ state \cite{yerokhin:24:sese}.
It is important to note that the numerical errors for $1s$ and
$2s$ states are not correlated, which can be seen from
the fact that individual SESE contributions in Table~\ref{tab:numres}
have different magnitudes for $1s$ and $2s$ states.
The agreement with the $\Za$-expansion results observed for the
normalized $2s$-$1s$ difference
not only confirms the consistency of the two methods but
also
serves as an independent confirmation that the numerical
uncertainties of our computations are under control.

\begin{figure*}
\centerline{
\resizebox{0.75\textwidth}{!}{%
  \includegraphics{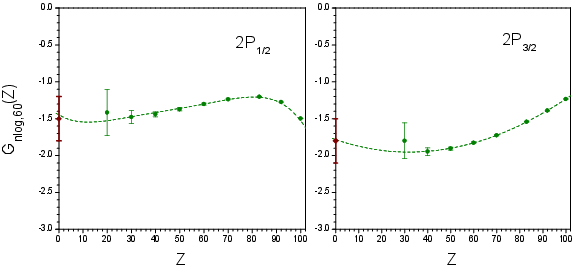}
}}
 \caption{The nonlogarithmic higher-order remainder
 $G_{\rm 60, nlog}$,
 see Eq.~(\ref{eq:Gnlog}) for definition,
 for the
 $2p_{1/2}$ state (left) and $2p_{3/2}$ state (right).
 Green dots
 denote our all-order numerical results, the brown dots
 at $Z = 0$ denote the $\Za$-expansion limiting value,
 the dashed line is a fit to guide the eye.
 \label{fig:2p}}
\end{figure*}

In Fig.~\ref{fig:2p} we present our all-order results for the
nonlogarithmic higher-order remainder $ G_{60,\rm nlog}$ for the
$2p_{1/2}$ and $2p_{3/2}$ states, in comparison with
the $\Za$-expansion results for the $B_{60}$ coefficient.
We conclude that  our nonpertibative results for the $2p$ states are
consistent with the $\Za$-expansion predictions, although  the
numerical accuracy is not yet sufficient for an independent verification
of the $\Za$-expansion result for the $B_{60}$ coefficient.

\section{Conclusion and outlook}

In this work 
we generalized the method for accelerating convergence of the
partial-wave expansion suggested in Ref.~\cite{sapirstein:23} from the
one-loop self-energy to the two-loop case and made it applicable 
for excited states.
We performed extensive calculations 
of the two-loop self-energy correction
for the $1s$, $2s$,
$2p_{1/2}$, and $2p_{3/2}$ states of hydrogen-like ions, with an
improved numerical accuracy and for a wider range of nuclear charges
than previously possible.
Accurate calculations of the two-loop QED effects for the
excited $n=2$ states
are essential to match the theoretical accuracy with the precision achieved in experimental studies
of the $2p_j$-$2s$ transitions in Li-like ions 
\cite{beiersdorfer:98,bosselmann:99,feili:00,beiersdorfer:05,epp:07,lestinsky:08}.

Our numerical all-order results are compared with those obtained
within the $\Za$ expansion. As found in Ref.~\cite{yerokhin:24:sese},
for the $1s$ state of hydrogen,
our nonperturbative results are in $3.5\,\sigma$ disagreement with the
$\Za$-expansion prediction.
In contrast, we find good agreement with the $\Za$ expansion for 
the normalized $2s$-$1s$ difference and the $2p_j$ states. 
This suggests that the discrepancy for the $1s$ state may stem 
from an additional "state-independent" contribution of order 
$\alpha^2(\Za)^6$ missing in the $\Za$-expansion calculations.
The state-independent
contributions are proportional to the
expectation value of the Dirac $\delta$ function, $\lbr \delta^3(r) \rbr$,
and vanish in the normalized $2s$-$1s$ difference as well as for states with $l>0$.

It is worth noting that the calculation in Ref.~\cite{pachucki:03:prl} for the $B_{60}$ coefficient of the $\Za$ expansion was incomplete, leaving some state-independent contributions unaccounted for. Therefore, a larger-than-expected missing contribution to the $B_{60}$ coefficient could be in principle responsible for the observed discrepancy.

In future, the developed method of the convergence acceleration of the partial-wave
expansion can be applied to calculations of
other two-loop QED effects, in particularly
to the two-loop self-energy correction
to the bound-electron $g$ factor.
First numerical results recently reported for this correction
\cite{sikora:24:tobe} indicated that their numerical accuracy
was severely limited by the convergence of the partial-wave expansion.
An implementation of the convergence-acceleration method
might open a way to extending these calculations to lower-$Z$ region.

\section{Acknowledgements}

Stimulating discussions with K.~Pachucki are gratefully acknowledged.
Intensive numerical computations reported in this work we carried out with
help of computer cluster of Max Planck Institute for Nuclear Physics.

\appendix

\section{Connection to previous definitions}

Below we give the exact relation between $E_{M'}$, $E_{P'}$, $E_{F'}$ and
$E_{M}$, $E_{P}$, $E_{F}$ used in previous studies
\cite{yerokhin:03:prl,yerokhin:05:sese,yerokhin:06:prl,yerokhin:09:sese,yerokhin:18:sese}:
\begin{align}
E_{M'} =&\ E_M - 2\,E_{o110} - E_{o020} - E_{o101}  - E_{n020} - E_{N3,P'}\,,
\\
E_{P'} = &\ E_P - 2\,E_{o200} - 2\,E_{n110} - 2\,E_{n200} -  E_{n101} + E_{N3,P'}\,,
\\
E_{F'} =&\ E_F + 2\,E_{o110} + E_{o020} + E_{o101} + 2\,E_{o200}
 \nonumber \\ &
  + 2\,E_{n110} + 2\,E_{n200} + E_{n020} + E_{n101}\,.
\end{align}


\end{document}